\documentstyle[11pt,epsf]{article}
\begin{document}

\title{\bf{Phenomenological Estimation of Parameters \\
in minimal Supergravity Model}}

\author{Jun Tabei\footnote{jun@hep.phys.waseda.ac.jp} 
and Hiroshi Hotta\footnote{hotta@hep.phys.waseda.ac.jp} \\
Department of Physics, Waseda University, Tokyo 169, Japan }

\maketitle
\begin{abstract}
Several parameters of the minimal supergravity model are estimated 
by the method of the renormalization group. 
In this model, five arbitrary parameters
$(m_0, m_{1/2}, \tan{\beta}, A_0, \mbox{sig}(\mu))$ are contained. 
$\tan{\beta}$ is evaluated as 7.5 by the likelihood analysis of 
the gauge and Yukawa coupling constants. 
Further, the trilinear coupling constant $A_0$ is fixed 
by the equation of $B_0=(A_0-1)m_0$ 
at the GUT scale $M_X \simeq 2\times10^{16}$(GeV) and by the 
Higgs potential at the Z-boson mass scale $M_Z=$91(GeV). 
As the results, (1) allowed $m_0-m_{1/2}$ regions, 
(2) the sparticle mass spectra, and 
(3) the lightest supersymmetric 
particle mass are shown in this paper. 
\end{abstract}

\section{Introduction}

In the minimal supergravity model(mSUGRA)\cite{Nilles}, 
five arbitrary parameters
($m_0$, $m_{1/2}$, $\tan{\beta}$, $A_0$, sig($\mu$))
are included. 
If these parameters are specified, all the mass of the particles 
are derived by the analysis of renormalization group flow. 
Recently, the mass of Standard model(SM) Higgs boson has been 
evaluated by LEP\cite{Erler}. 
The reported Higgs mass is slight smaller than expected. 
This fact encourages the mSUGRA, because such light value 
of the Higgs mass is supported as one of the features of 
this model. 
If the mass of the lightest neutral Higgs boson in mSUGRA 
is close to the mentioned above the mass of Higgs boson 
in the SM\cite{Okada}, 
a constraint condition among $m_0$, $m_{1/2}$ and $\tan{\beta}$ 
is obtained. 
In order to determine their values completely, some additional 
conditions are introduced as follows: 

\begin{itemize}

\item[i)] Assuming the SU(5) GUT scenario, 
the gauge and 
the Yukawa coupling constants of D-type quarks and leptons are 
unified at GUT scale $M_X$, respectively\cite{Barger}. 

\item[ii)] As one of the conditions of mSUGRA, 
$B_0$ is equal to $(A_0-1)m_0$ at $M_X$\cite{Nilles}. 
At $M_Z$ scale, two different $B(M_{Z})$ values are obtained; 
(a) by RGE development from $M_X$ scale, and 
(b) by the Higgs potential at $M_Z$ scale directly. 
$A_0$ is tuned numerically to make both $B(M_{Z})$ values match 
from each other. 

\item[iii)] The condition sig$(\mu)>0$ is derived 
from the analysis of the g-2 experiment\cite{Chattopadhyay}. 
\end{itemize}
As the results, $m_0-m_{1/2}$ regions, mass spectra, and 
the lightest supersymmetric particle(LSP) is 
shown in the following sections by making use of 
the above conditions in this paper. 

\section{Analysis of Parameters}
\subsection{On the $\tan{\beta}$}
\noindent
On the optimization of $\tan{\beta}$ value is discussed by 
the two-loop level renormalization group equations(RGEs) 
\cite{Barger,Castano}
of the gauge and Yukawa 
coupling constants in this section. 
Boundary conditions of the RGEs are given 
at Z-boson mass scale $M_Z$. 
Their values at $M_Z$ scale are referred 
from the latest Particle Data Group\cite{Particle}. 
\begin{itemize}

\item[1)] gauge coupling constants \\
The average value $\alpha_X$ among 
$\alpha_1,\alpha_2$ and $\alpha_3$ is introduced 
around GUT scale $M_X$. 
$\chi^2$ fitting function is made of the squared sum of 
each differences between $\alpha_X$ and 
$\alpha_i(i=1 \sim 3)$ as follows: 

\begin{equation}
\chi^2=\sum_{i=1}^3\left(\frac{\alpha_i-\alpha_X}{\sigma_i}\right)^2
\end{equation}

where, $\sigma_i$ is the standard deviation of $\alpha_i$.
The energy scale where the $\chi^2$ becomes minimal 
is defined as $M_X$ value. 
The value of $\chi^2$ is shown as a function of 
$\tan{\beta}$ in fig.1 . 
By the way, another mass scale parameter 
$M_{SUSY} \simeq 1$(TeV) 
is used to be introduced as a turning-on point of the sparticles' 
effect, however, since $a_3$ is estimated 0.118 by the latest 
P.D.G., such $M_{SUSY}$ becomes 
less than the scale $M_Z$\cite{Amaldi}. 
Therefore, the $M_{SUSY}$ is neglected in this paper. 

\begin{figure}
   \epsfxsize= 10 cm
   \centerline{\epsfbox{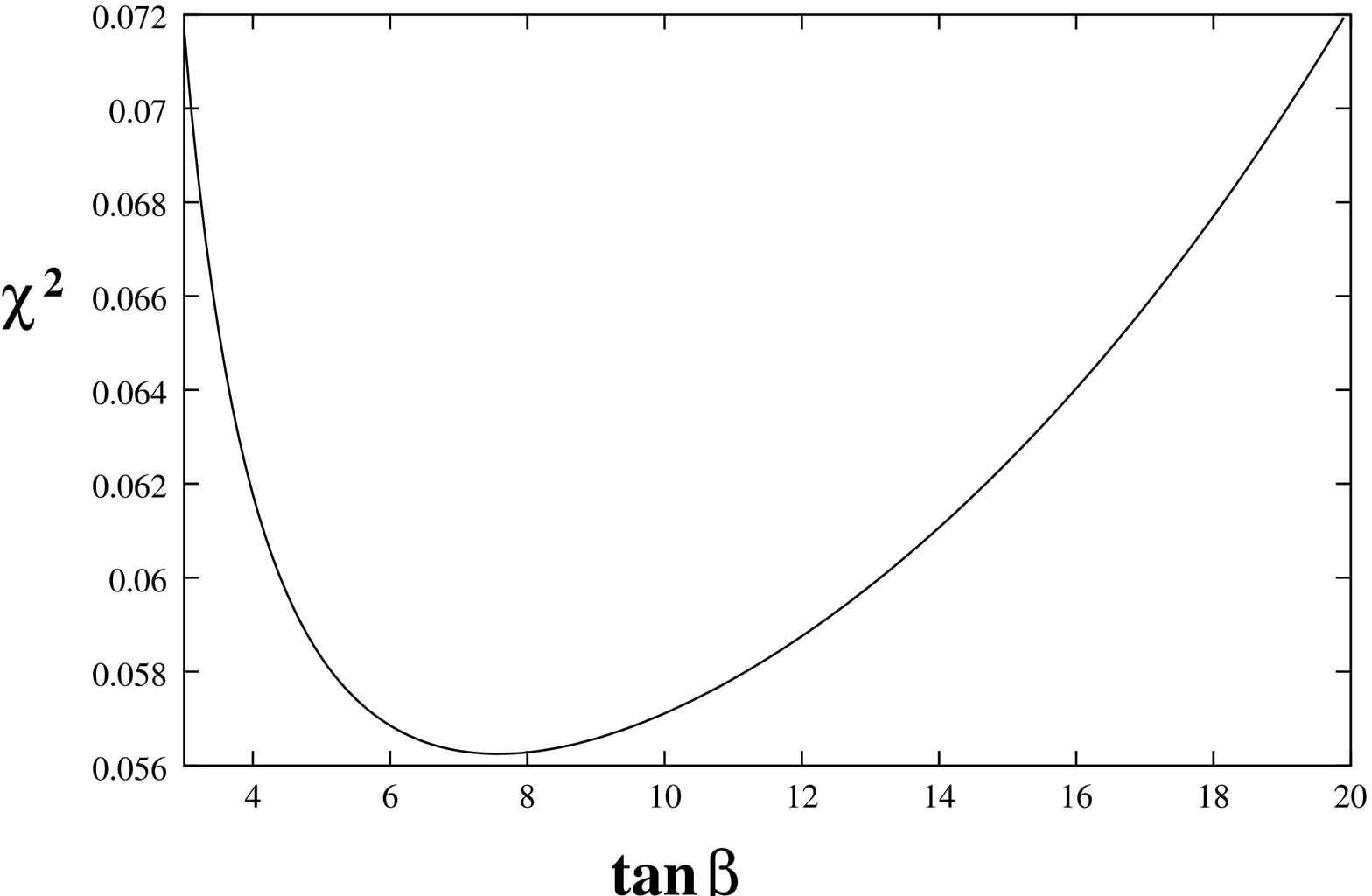}}
\caption{
}
\end{figure}

\item[2)] Yukawa coupling constants \\
When the SU(5) scenario is presumed \cite{Barger}, 
Yukawa coupling constants of D-type quark and lepton 
is supposed to coincide at $M_X$.
Their $\chi^2$ fitting function 
is defined as the following equations: 

\begin{equation}
\chi^2=\sum_{{i=1,2,3}\atop{q=D,L}}\left(
\frac{\alpha_{q_i}-\alpha_{X_i}}{\alpha_{X_i}}
\right)^2
\end{equation}

\begin{equation}
\alpha_{X_i}=\left(
\alpha_{D_i}+\alpha_{L_i}
\right)/2
\end{equation}

where, the suffix $i$ denotes the generation, 
and $q$ implies D-type quarks or Leptons. 
Moreover, with taking the CKM matrix\cite{Particle} 
into consideration, 
Yukawa coupling constants of RGEs are diagonalized. 
The $\tan{\beta}$ dependence on $\chi^2$ at $M_X$ scale 
is shown in fig. 2. 
In the minimal SO (10) scenario \cite{Barger}, 
once the mass spectra of the particles 
in the lower energy scale 
region are fixed to their observed values, required 
unification of Yukawa coupling constants is failed. 
With likelihood analysis of $\tan{\beta}$, 
it is failed to determine 
the optimized value of $\tan{\beta}$ to unify the Yukawa 
coupling constants.
\end{itemize}

\begin{figure}
   \epsfxsize= 10 cm
   \centerline{\epsfbox{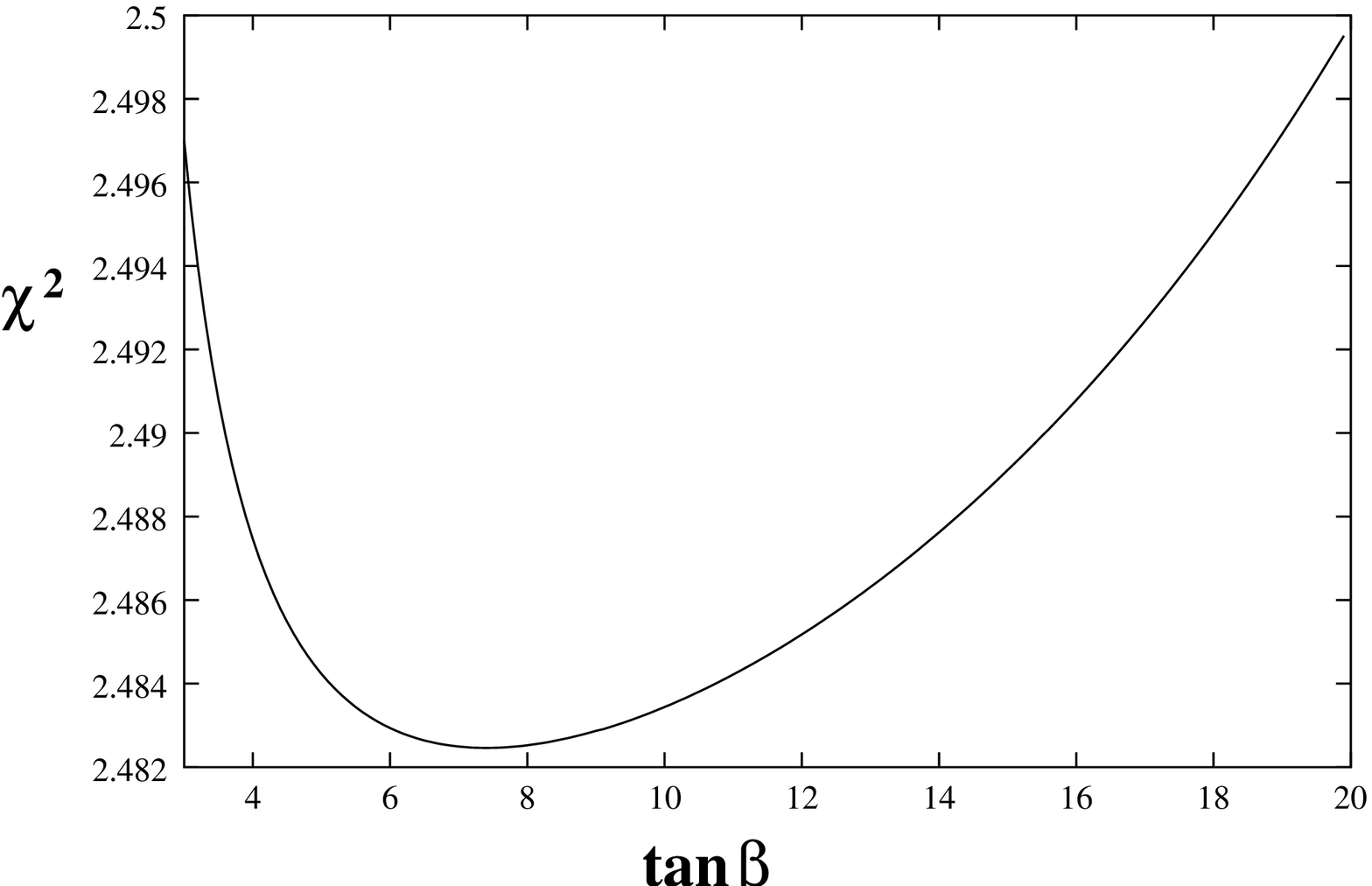}}
\caption{
}
\end{figure}
\noindent
According to the above mentioned 1) and 2), 
the value of $\tan{\beta}$ is implied as much as 7.5 . 
As the result, in order to keep the unification among gauge 
coupling constants, 
it is impossible to unify the Yukawa coupling constants 
at $M_X$ as shown in fig.3 and fig.4 . 

\begin{figure}
   \epsfxsize= 10 cm
   \centerline{\epsfbox{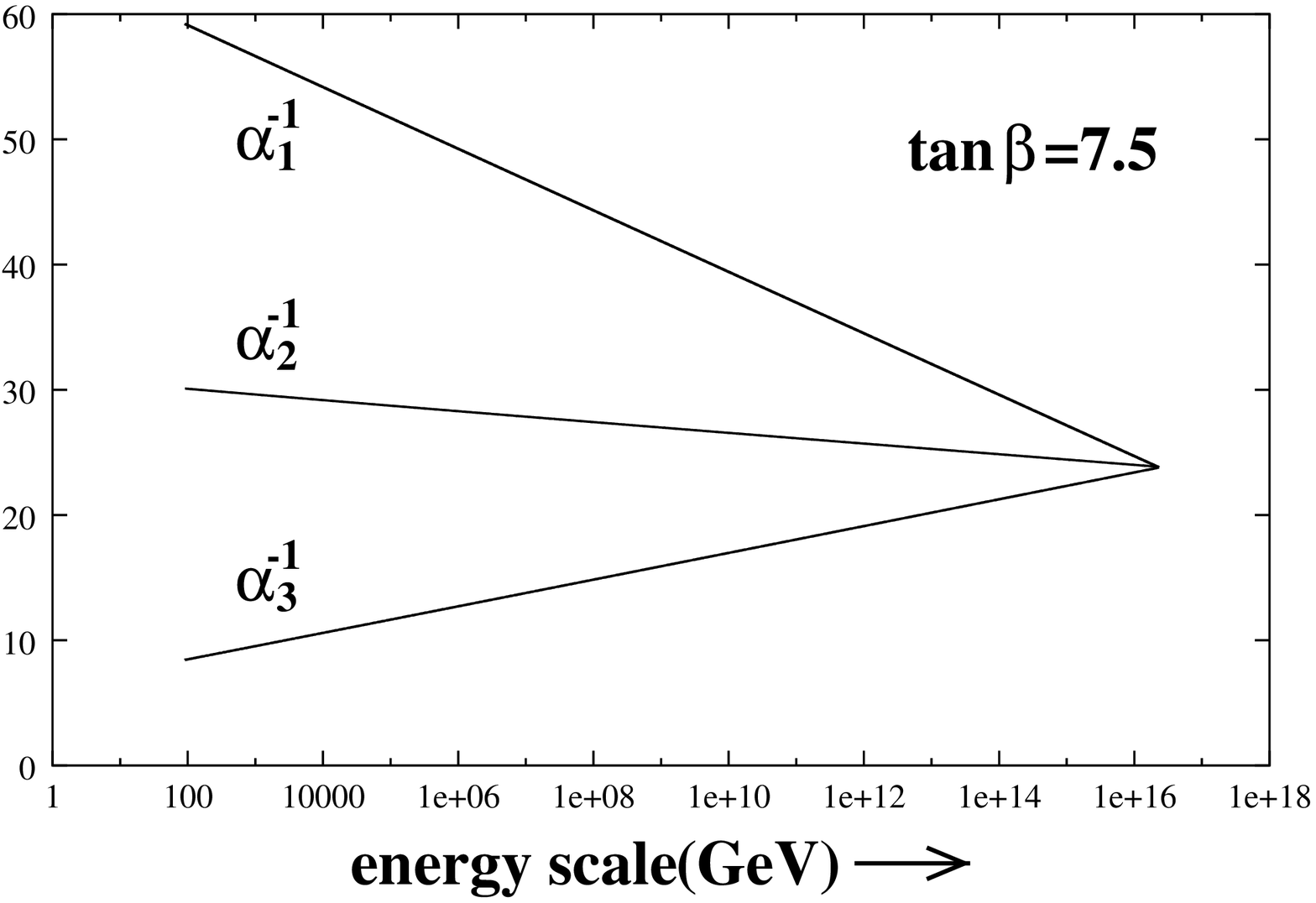}}
\caption{
}
\end{figure}
\begin{figure}
   \epsfxsize= 10 cm
   \centerline{\epsfbox{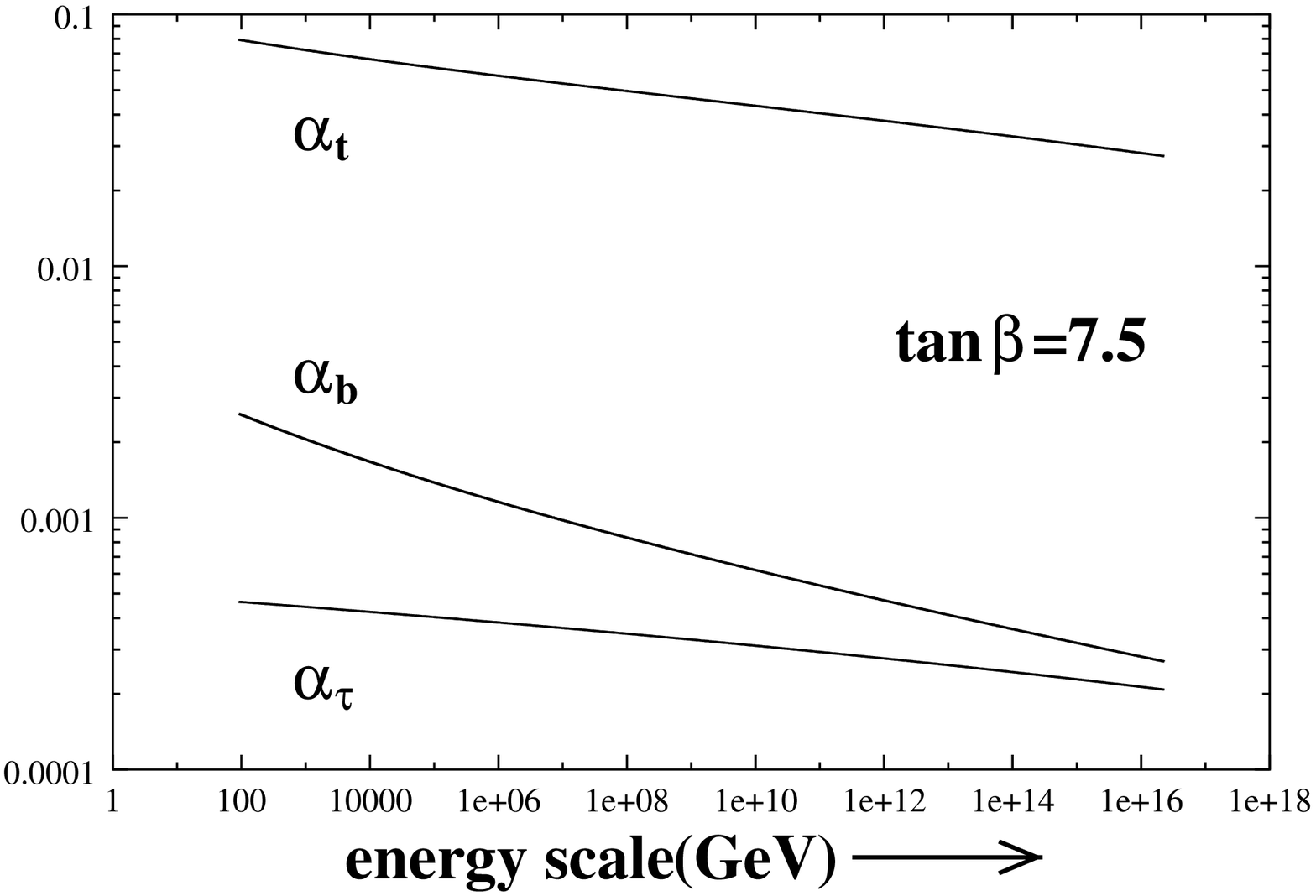}}
\caption{
}
\end{figure}

\subsection{On the $A_0$}
\noindent
As mentioned in avant, one $B(M_{Z})$ value is obtained by 
1-loop level RGEs \cite{Castano,Inoue} 
development with boundary 
constraint condition $B_0=(A_0-1)m_0$ at $M_X$ scale. 
Another $B(M_{Z})$ is calculated at $M_Z$ as 

\begin{equation}
\mu^2=-\frac{m_Z^2}{2}+
\frac{m_{h_d}^2-m_{h_u}^2\tan^2{\beta}}{\tan^2{\beta}}
\end{equation}
\begin{equation}
B(M_{Z})=-\frac{1}{2\mu}\left(
m_{h_d}^2-m_{h_u}^2+2\mu^2\right)
\sin{2\beta}
\end{equation}
where, $m_{h_d}$ and $m_{h_u}$ are soft breaking mass parameters of 
two Higgs doublet at $M_Z$. 
$A_0$ is determined by the numerical tuning 
to coincide both $B(M_{Z})$ values. 
It is shown that $A_0$ at 
$m_{1/2}/m_0$=$8/3$, 1, $3/8$ as functions of $\tan{\beta}$
in fig.5 . 
\begin{figure}
   \epsfxsize= 10 cm
   \centerline{\epsfbox{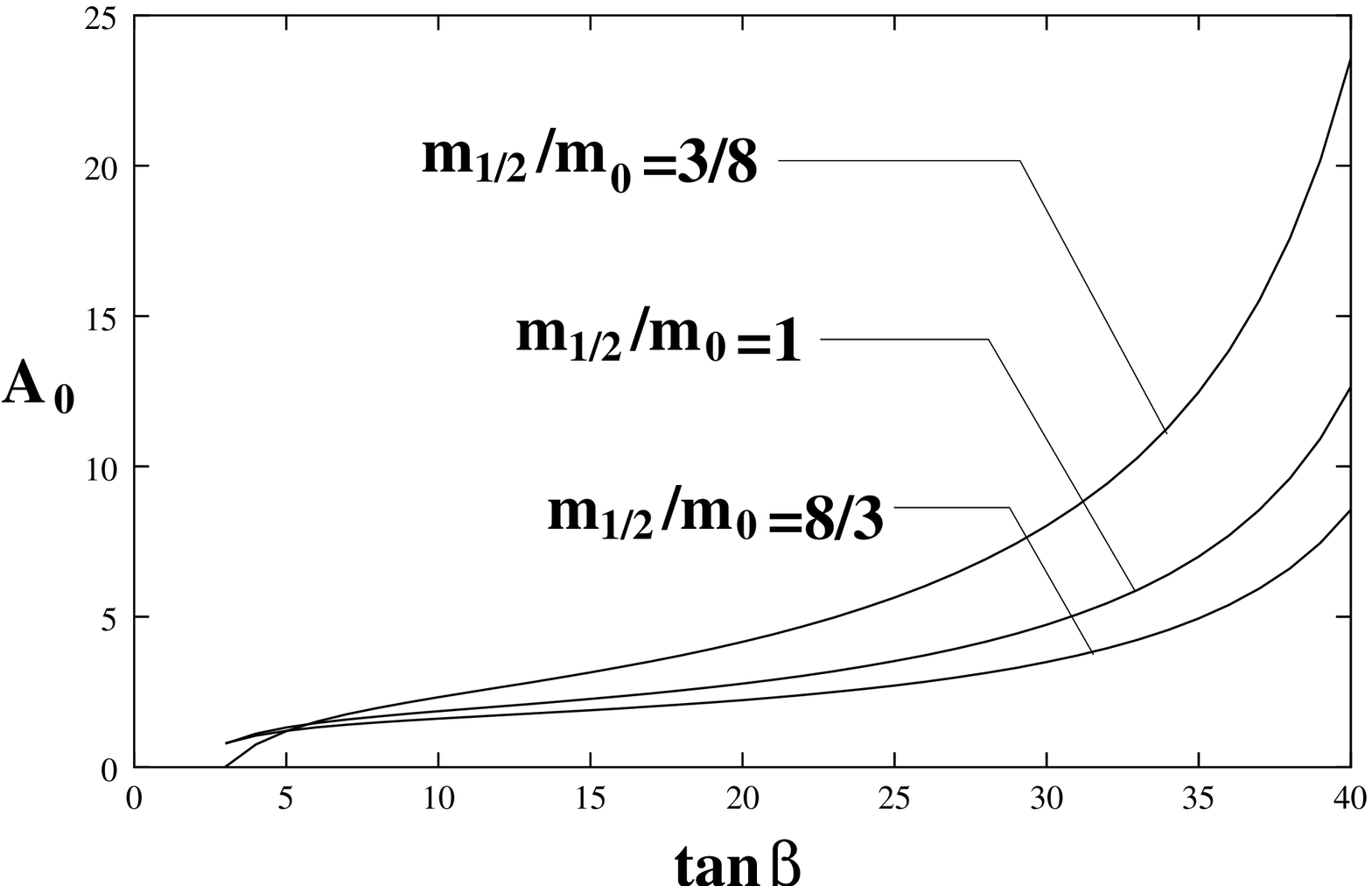}}
\caption{
}
\end{figure}

\subsection{On the sig($\mu$)}
\noindent
sig$(\mu)>0$ is assumed by the result of 
the g-2 experiment\cite{Chattopadhyay}.

\subsection{On the experiment result of LEP.}
\noindent
As the experimental results of LEP \cite{LEP}, 
the lightest neutral Higgs mass in the SUSY models 
exists in the region from 90(GeV) to 130(GeV). 
However, the lightest neutral Higgs mass is supposed 
to be beyond 115(GeV) like SM Higgs \cite{Erler} in this paper, 
because such small mass as 90(GeV) makes difficulties to 
evaluate $m_0$ and $m_{1/2}$.

\section{Numerical results}
\begin{figure}
   \epsfxsize= 10 cm
   \centerline{\epsfbox{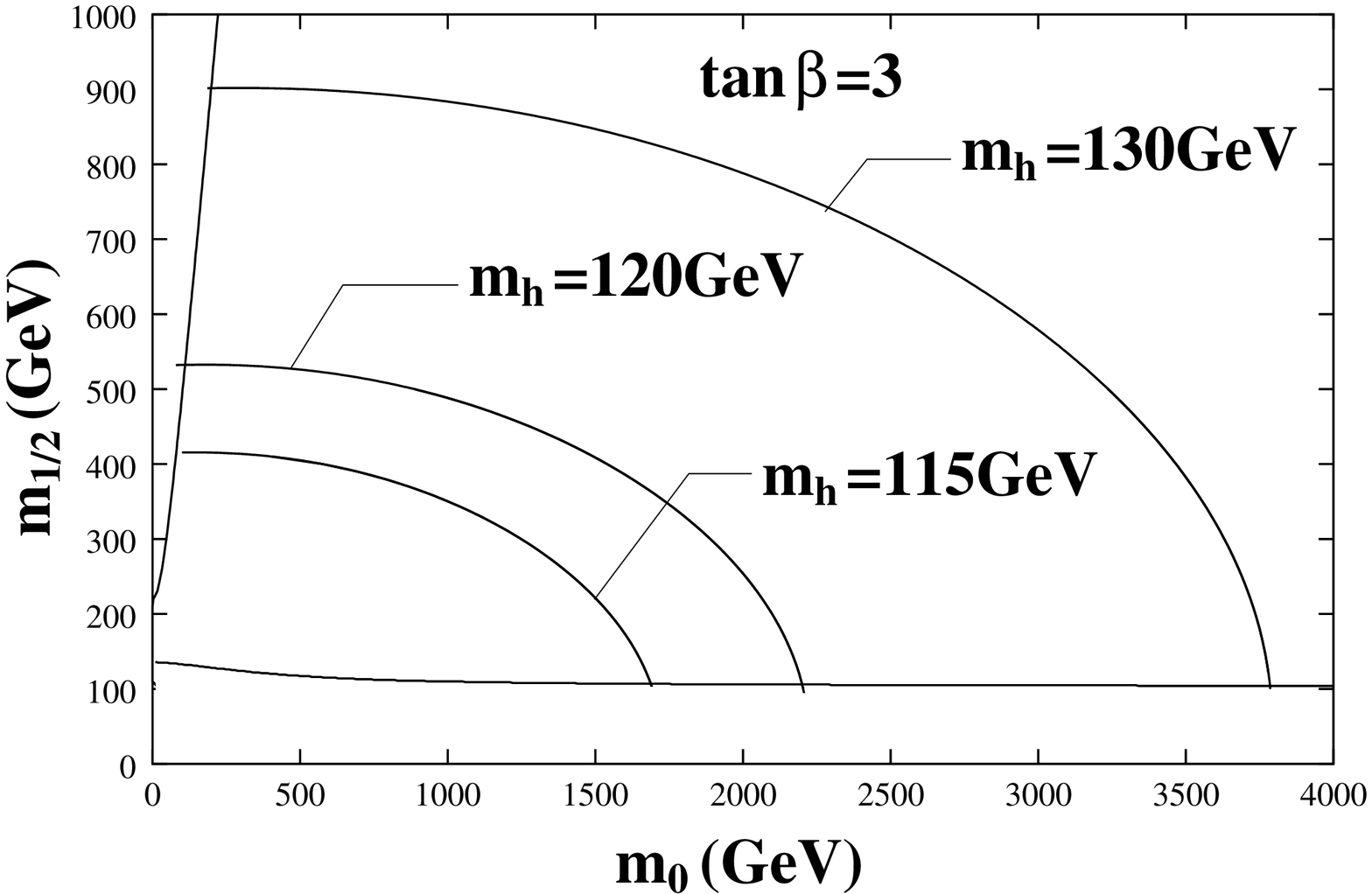}}
\caption{
}
\end{figure}

\begin{figure}
   \epsfxsize= 10 cm
   \centerline{\epsfbox{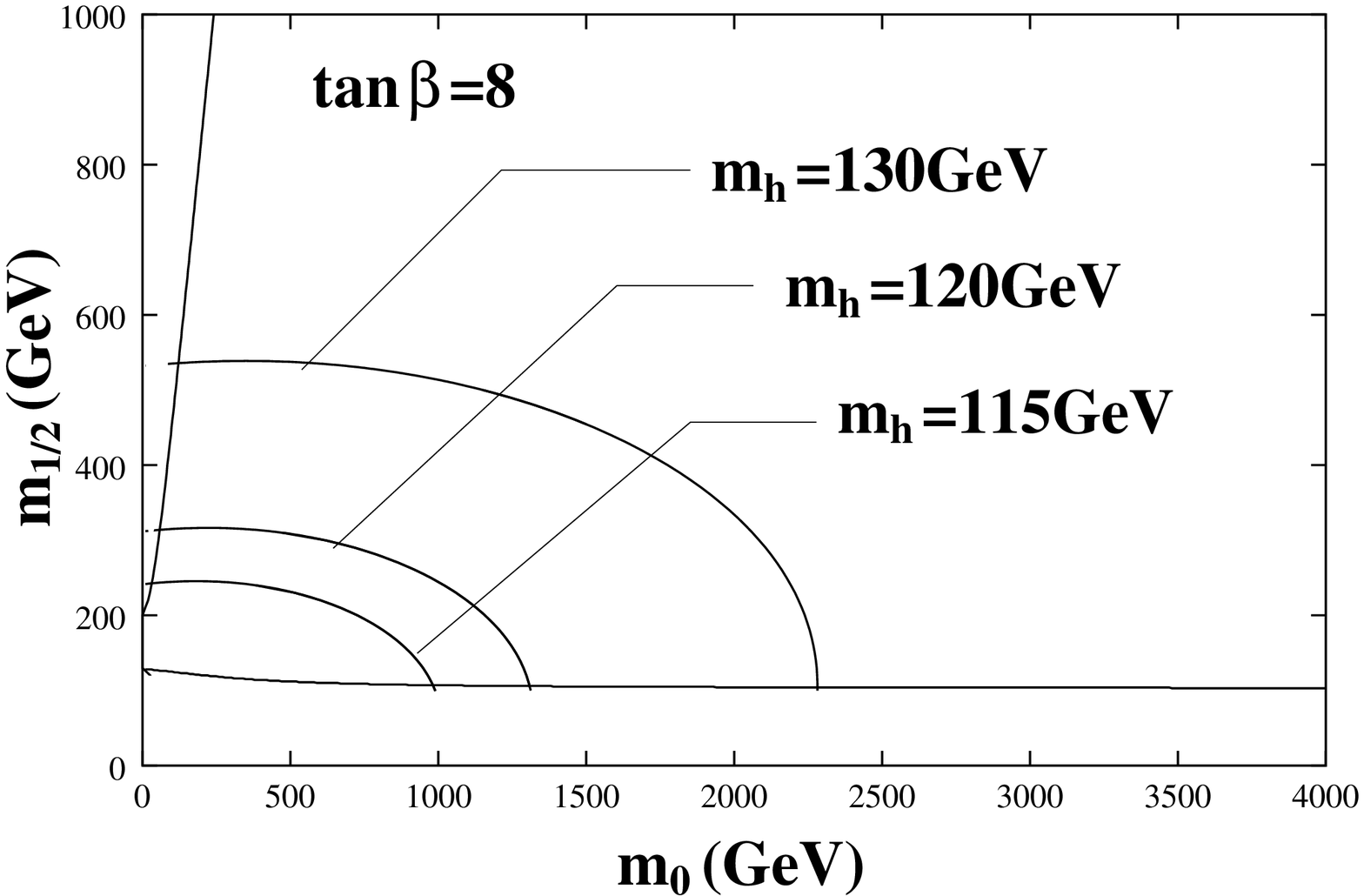}}
\caption{
}
\end{figure}

\begin{figure}
   \epsfxsize= 10 cm
   \centerline{\epsfbox{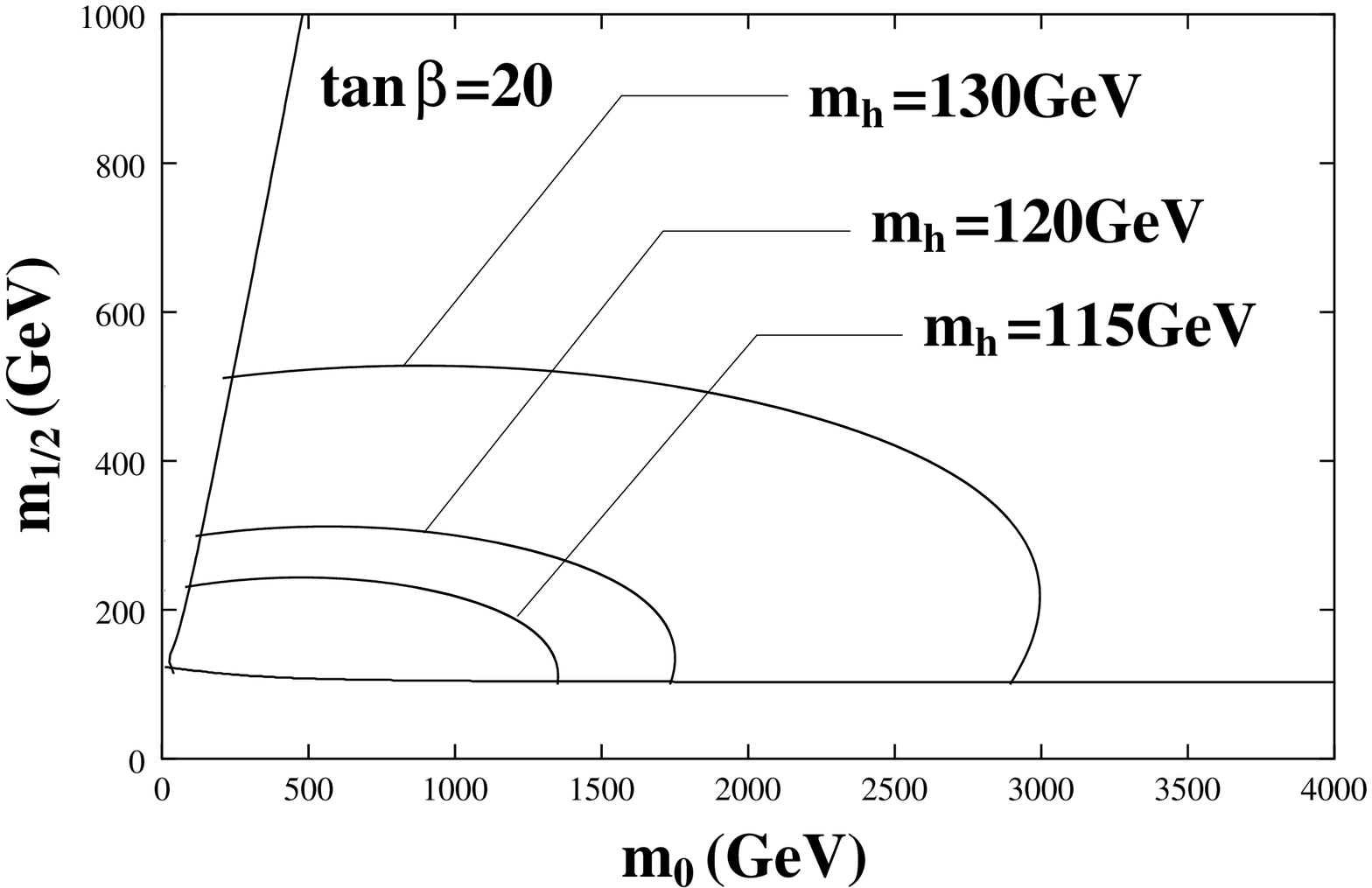}}
\caption{
}
\end{figure}
\begin{figure}
   \epsfxsize= 10 cm
   \centerline{\epsfbox{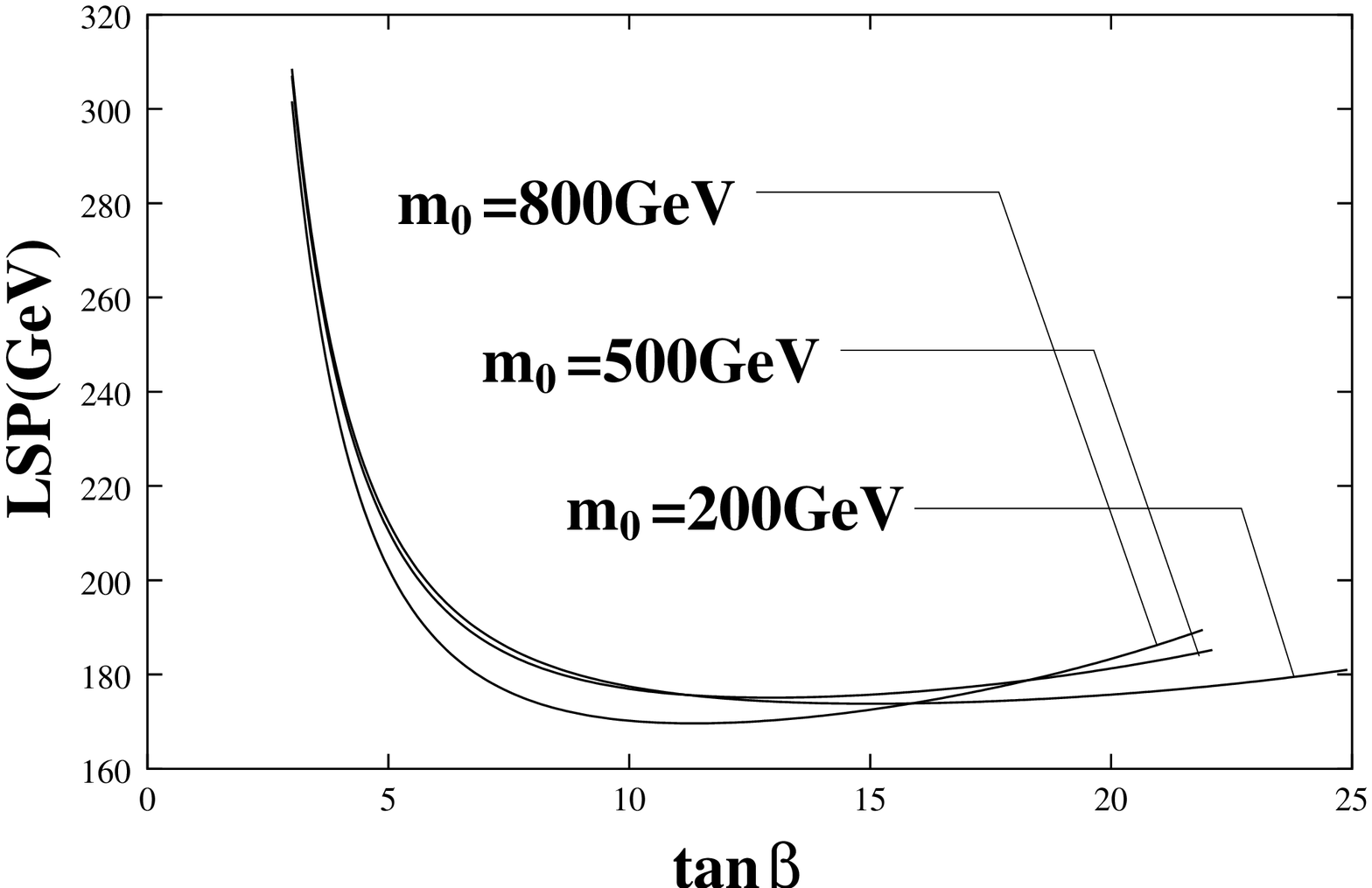}}
\caption{
}
\end{figure}

\begin{itemize}

\item[i)] $m_0-m_{1/2}$ plane 

Higgs mass is calculated by 
the 1-loop level RGE analysis as follows \cite{Carena}: 

\begin{equation}
m_h=m_{h,(tree)}+\Delta m_{h,(1-loop)}
\end{equation}

Corresponding $m_0-m_{1/2}$ regions 
are shown with several Higgs mass settings(115,120 and 130 (GeV)) 
with $\tan{\beta}$=3, 8, and 20 in fig.5, fig.6, and fig.7, 
respectively. 

The region below the horizontal line is excluded by 
the chargino mass bound $m_{\chi^\pm}>91(GeV)$.
Moreover, the region in the left side of the ordinate line 
is excluded by the fact that it is not 
charged particle(tau slepton), the LSP 
must be neutral particle(neutralino). 

\item[ii)] mass spectra 

$m_0$ is assumed to coincide with the result of g-2 experiment. 
The predictions of the mass spectra with several 
different $\tan{\beta}$ are listed on Table 1. 

\begin{table}
\caption{
Mass Spectra of mSUGRA(unit is GeV). }
\begin{center}
\begin{tabular}{ l l l l }
\hline
\hline
 $\tan{\beta}$(non-dimensional)  &3      &8      &20 \\
 $m_0$  &300    &300    &600 \\
 $m_h$  &120    &120    &120 \\
\hline
 $m_0$  &531    &316    &316 \\
 $\mu$  &1031   &626    &803 \\
 B      &-416   &-99    &-34 \\
 $A_0$(non-dimensional)  &0.473  &1.70   &2.37 \\
 heavier neutral Higgs   &1197   &709    &744 \\
 CP-odd Higgs   &1196   &709    &744 \\
 Charged Higgs          &1198   &713    &748 \\
 chargino1      &461    &271    &276 \\
 chargino2      &1041   &640    &813 \\
 neutralino1    &237    &140    &141 \\
 neutralino2    &461    &271    &276 \\
 neutralino3    &-1032  &-630   &-807 \\
 neutralino4    &1042   &640    &812 \\
\hline
 $m_{\tilde{u}_1}$      &1391   &861    &1006 \\
 $m_{\tilde{c}_1}$      &1391   &861    &1006 \\
 $m_{\tilde{u}_2}$      &1433   &884    &1026 \\
 $m_{\tilde{c}_2}$      &1433   &884    &1026 \\
 $m_{\tilde{t}_1}$      &1042   &580    &495 \\
 $m_{\tilde{t}_2}$      &1341   &840    &820 \\
 $m_{\tilde{d}_1}$      &1387   &859    &1005 \\
 $m_{\tilde{s}_1}$      &1387   &859    &1005 \\
 $m_{\tilde{d}_2}$      &1434   &888    &1029 \\
 $m_{\tilde{s}_2}$      &1387   &859    &1005 \\
 $m_{\tilde{b}_1}$      &1291   &770    &725 \\
 $m_{\tilde{b}_2}$      &1385   &850    &863 \\
\hline
 $m_{\tilde{e}_1}$      &366    &327    &614 \\
 $m_{\tilde{\mu}_1}$    &366    &327    &614 \\
 $m_{\tilde{e}_2}$      &481    &369    &637 \\
 $m_{\tilde{\mu}_2}$    &481    &369    &637 \\
 $m_{\tilde{\tau}_1}$   &365    &321    &515 \\
 $m_{\tilde{\tau}_2}$   &481    &369    &637 \\
 $m_{\tilde{\nu}_e}$    &482    &371    &638 \\
 $m_{\tilde{\nu}_\mu}$  &482    &371    &638 \\
 $m_{\tilde{\nu}_\tau}$ &482    &371    &638 \\
\hline
\hline
\end{tabular}
\end{center}
\end{table}

\item[iii)] lightest supersymmetric particle(LSP) 
The mass of the LSP
as a function of $\tan{\beta}$ is shown 
at $m_0$=200, 500 and 800 (GeV) in fig. 8.
Note that the neutralino is most plausible candidate 
of the LSP. 
\end{itemize}

\section{Summary and Conclusion}

The parameters of mSUGRA are estimated 
by several conditions phenomenologically. 
$\tan{\beta}$ is estimated about 7.5 by using RGEs of the 
gauge and Yukawa coupling constants. 
$A_0$ is fixed by the constraint at $M_X$ and 
the conditions at $M_Z$. 
The allowed $m_0-m_{1/2}$ regions are shown with 
several values of 
the Higgs boson's mass. 
The mass spectra and the mass of the LSP 
are calculated. 
Needless-to-say, it is important to find at least 
one supersymmetric particle experimentally 
for the reality of mSUGRA or other supersymmetric models. 
With the results shown in this paper, 
the supersymmetric particles are not so much 
heavy, and possible to be observed by colliders 
in the near future.

\end{document}